\documentclass[prd,a4paper,twocolumn,nofootinbib,aps,10pt,preprintnumbers,longbibliography]{revtex4-2}

\usepackage[colorlinks=true,linkcolor=black,citecolor=blue,urlcolor=blue, pdfborder={0 0 0}]{hyperref}
 \usepackage{amssymb,amsmath,amsthm,cancel,graphicx}
\usepackage{mathtools}
\usepackage[utf8]{inputenc}
\usepackage{color}
\usepackage[table,xcdraw,dvipsnames]{xcolor}
\usepackage{multirow}
\usepackage[normalem]{ulem}    
\usepackage{bm} 
\usepackage{slashed} 
\usepackage{booktabs}
\usepackage[table,xcdraw,dvipsnames]{xcolor}
\hypersetup{
	bookmarks=true,         
	unicode=true,          
	pdftoolbar=true,        
	pdfmenubar=true,        
	pdffitwindow=false,     
	pdfstartview={FitH},    
	pdftitle={NS and DM},    
	pdfauthor={Author},     
	pdfsubject={Subject},   
	pdfcreator={Creator},   
	pdfproducer={Producer}, 
	pdfkeywords={keyword1} {key2} {key3}, 
	pdfnewwindow=true,      
	colorlinks=true,       
	linkcolor=black,          
	citecolor=black,        
	filecolor=black,      
	urlcolor=black           
}
\usepackage{physics}

\def\eq#1{{Eq.~(\ref{#1})}}
\def\eqs#1#2{{Eqs.~(\ref{#1})-(\ref{#2})}}

\def\diag{{\rm diag}}

\def\mL{\mathcal{L}}
\def\mM{\mathcal{M}}
\def\mN{\mathcal{N}}

\def\subtopic#1{{\noindent\textbf{\emph{#1}\ }}}

\def\diag{{\rm diag}}

\newcommand{\be}{\begin{equation}}
\newcommand{\ee}{\end{equation}}
\newcommand{\bea}{\begin{eqnarray}}
\newcommand{\eea}{\end{eqnarray}}

\newcommand{\W}{\scriptscriptstyle{W}}

\newcommand{\mWW}{m_{\scriptscriptstyle{W\!W}}}

\newcommand{\SM}{{\rm SM}}
\newcommand{\MW}{M_W}
\newcommand{\MZ}{M_Z}

\newcommand{\sW}{\sin \theta_W}

\newcommand{\ssW}{s_{\W}}

\newcommand{\AB}{\scriptscriptstyle{AB}}
\newcommand{\LL}{\scriptscriptstyle{LL}}
\newcommand{\TT}{\scriptscriptstyle{TT}}
\newcommand{\LT}{\scriptscriptstyle{LT}}

\newcommand{\Ct}{c_{\Theta}}
\newcommand{\St}{s_{\Theta}}

\newcommand{\WW}{\scriptscriptstyle{W\!W}}

\newcommand{\betaW}{\beta_{\W}}
\newcommand*\xbar[1]{%
  \hbox{\;%
    \vbox{%
      \hrule height 0.5pt 
      \kern0.5ex
      \hbox{%
        \kern-0.25em
        \ensuremath{#1}%
        \kern-0.07em
      }%
    }%
  }%
} 

\definecolor{KBFIred}{RGB}{163,35,47}

\begin{document}
	
\title{\Large \textbf{\color{KBFIred}{Testing the unitarity of the light neutrino mixing matrix}}}

\vspace{1cm}
\author{\bf E. Gabrielli$^{{a,b,c}}$} 
\author{\bf A. Lind$^{{d}}$}
\author{\bf L. Marzola$^{{c,e}}$}
\author{\bf K. M\"u\"ursepp$^{{c,d}}$}
\author{\bf E. Nardi$^{{c,d}}$}

\vspace{0.5cm}

\affiliation{$^{a}$ Physics Department, University of Trieste, Strada Costiera 11, I-34151 Trieste, Italy}
\affiliation{$^{b}$ INFN, Sezione di Trieste, Via Valerio 2, I-34127 Trieste, Italy}
\affiliation{$^{c}$ Laboratory of High-Energy and Computational Physics, NICPB, R\"avala 10,  10143 Tallinn, Estonia}
\affiliation{$^{d}$ INFN, Laboratori Nazionali di Frascati, C.P. 13, 100044 Frascati, Italy  }
\affiliation{$^{e}$ Institute of Computer Science, University of Tartu, Narva mnt 18, 51009 Tartu, Estonia}

\begin{abstract}
We propose a novel test of the unitarity of the Pontecorvo–Maki–Nakagawa–Sakata (PMNS) mixing matrix at collider experiments. Our approach exploits the incomplete cancellation between $t$-channel neutrino exchange and $s$-channel gauge-boson contributions that arises in the presence of violation of the flavor-diagonal PMNS unitarity conditions in weak boson pair production, leading to an anomalous growth of the cross section with energy. Such effects are generic in extensions of the Standard Model in which light neutrinos mix with heavier states, and can manifest at colliders as long as the characteristic energy of the process remains below the mass threshold of the new degrees of freedom. After briefly reviewing these scenarios, we employ our strategy to derive model-independent bounds on flavor diagonal unitarity-violating effects using LEP~II data. We then present sensitivity projections for future lepton and hadron colliders, demonstrating that they are well suited to probe the unitarity of the neutrino mixing matrix with this method.
\end{abstract}
		\vspace{1cm}
		\maketitle

\section{Introduction}

In order to explain the suppression of neutrino masses, 
the Standard Model (SM) is often extended with additional neutral fermions generally assumed to be heavier than the 
electroweak scale. In a theory presenting $n$ additional states on top of the three SM neutrinos, the mass matrix of the neutral fermions has  the following structure:
\begin{equation}
 \mM = \begin{pmatrix}
      \mM_{33} & \mM_{3n} \\
      \mM_{n3} & \mM_{nn}
  \end{pmatrix}   \;,
 \label{eq:M}
\end{equation}
where $\mM_{33}$ and $\mM_{nn}$ are symmetric matrices, and $\mM_{n3}= \mM_{3n}^T$.  
The complex-symmetric matrix $\mM$ can be expressed via Takagi  factorization as $\mM = U^* \hat M U^\dagger$,  where $\hat M$ is a real diagonal matrix 
whose diagonal elements are the nonnegative square roots of the eigenvalues of 
 $ \mM^{\dagger} \mM =U \hat M^{2}U^{\dagger}$,  
and $U$ is a unitary matrix ($U^{\dagger}=U^{-1}$). The mass eigenvalues in $\hat M$ can be sorted into two sets: one containing the three masses of the light SM neutrinos, $\hat m_{\nu_i}$ with $i=1,\,2,\,3$, and one comprising the masses $\hat M_{\mN_n}$ of the $n$ additional heavy neutral fermions $\mN_n$. The unitary matrix $U$ can be written in block form as
\begin{equation}
U =  \begin{pmatrix}
      U_{\nu} & U_{3n} \\
      U_{n3} & U_{nn}
  \end{pmatrix}   \;,
 \label{eq:U}
\end{equation}
where $U_{\nu}$ describes the superposition of the light neutrinos mass eigenstates $\nu_i$ 
in the active neutrino state of flavor $\alpha$
\begin{equation}
\ket{\nu^\alpha} = \sum_{i=1}^3 (U_\nu^*)^{\alpha i} \ket{\nu_i} + \sum_{k=1}^n (U_{3n}^*)^{\alpha k}\ket{\mN_k}\,,
\end{equation}
and is commonly denoted as the Pontecorvo-Maki-Nakagawa-Sakata (PMNS)~\cite{Pontecorvo:1957qd,Maki:1962mu} 
matrix.\footnote{We work in the flavor basis, wherein the mass matrix 
of the charged leptons is diagonal with real nonnegative entries.} 
In the presence of mixing between the light neutrinos and the heavy neutral fermions, the SM leptonic charged current Lagrangian is modified as
\begin{equation}
    \mL = -\frac{g}{\sqrt{2}} W^-_\mu \bar \ell_\alpha \gamma^\mu 
    \left(U_{\nu}^{\alpha i} \nu_i + U_{3n}^{\alpha i} \mN_i\right) + H.c. 
\end{equation}
with $\ell_\alpha$, $\alpha= (e,\mu,\tau)$, being the vector of the left-handed (LH) SM charged leptons, and with $U_{3n}$ accounting for the new heavy $\mN_i$ components in $\nu_\alpha$. Imposing the unitarity of the matrix $U$ straightforwardly implies deviations from unitarity for the PMNS matrix:
\begin{eqnarray}
    \sum_{i=1}^3 |U_{\nu}^{\alpha i}|^2 &=& 1-  \sum_{i=4}^n |U_{3n}^{\alpha i}|^2 = 1-\delta_\alpha \,,
    \label{eq:diag}\\
    \sum_{i=1}^3 U_{\nu}^{\alpha i } {U_{\nu}^{\beta i\;*}} &=& -  \sum_{i=4}^n  U_{3n}^{\alpha i} {U_{3n}^{\beta i}}^*= \epsilon_{\alpha\beta} \quad\   (\beta \neq \alpha) \,.
      \label{eq:offdiag}
\end{eqnarray}
The parameters $\epsilon_{\alpha\beta}$ ($\alpha,\beta=e,\mu,\tau$) in \eq{eq:offdiag} are responsible for enhancing lepton flavor violating (LFV) transitions such as $\mu \to e\gamma$, $\tau \to \mu \gamma$ and $\tau \to e \gamma$, with respect to the SM rates. Experimental bounds on these rare processes are particularly strong, and provide constraints at the level of  $|\epsilon_{\mu e}|\lesssim 10^{-5}$ and 
$|\epsilon_{\tau \mu}|, \, |\epsilon_{\tau e}|\lesssim 6\times 10^{-3}$~\cite{Forero:2011pc,Basso:2013jka,Antusch:2014woa}. 
The parameters $\delta_\alpha\geq 0$ ($\alpha=e,\mu,\tau$) in \eq{eq:diag}  encode 
flavor diagonal non-unitarity effects. They are constrained 
by global fits to high precision electroweak data at the level 
of $\delta_e \lesssim 2\times 10^{-3} $,  $\delta_\mu \lesssim 2\times 10^{-4}$ and 
$\delta_\tau \lesssim 5\times 10^{-3} $~\cite{Nardi:1994iv,Antusch:2006vwa,Basso:2013jka,Antusch:2014woa},
while somewhat weaker bounds are also obtained from neutrino oscillations data~\cite{Parke:2015goa,Ellis:2020hus,Hu:2020oba,
Denton:2021mso,Kozynets:2024xgt}.

An interesting way to test diagonal non-unitarity effects sourced by the PMNS matrix is to search for processes in which such effects may spoil perturbative unitarity of certain cross sections in high energy collisions. A particularly promising environment is then provided by the lepton–antilepton colliders, and, to some degree, by the upcoming High-Luminosity (HL) phase of the LHC experiment. Indeed, as long as the energy of the process remains below the scale of the heavy neutral fermion masses, non-unitarity effects can induce an anomalous energy scaling of the cross section in interactions involving 
the exchange of the light neutrinos. The effect can then be investigated at hadron colliders via the $W^+ W^-$ fusion, and at lepton colliders via the $W^+ W^-$ production. The same underlying mechanism has been recently exploited 
to test the unitarity of the Cabibbo–Kobayashi–Maskawa quark mixing matrix, by studying the anomalous growth of the $pp \to W^+ W^-$ cross section at the LHC, and at various future hadron colliders \cite{Gabrielli:2024bjw} (see also \cite{Eboli:2025vks} for 
similar bounds on non-unitarity for the first row of the CKM matrix.)
Assessing how accurately the proposed method can constrain non-unitarity effects in the neutrino mixing with the available data, as well as with measurements at future $pp$, $e^+e^-$ and $\mu^+\mu^-$ colliders, is the main goal of this work. To this purpose, in the next section, we examine the theoretical expectations for the size of the parameters $\delta_\alpha$ in well-motivated neutrino mass models. Subsequently, we derive bounds on $\delta_\alpha$ from the available LEP data, as well as for the projected sensitivity of the future high-energy facilities such as HL-LHC, FCC-ee, FCC-hh, CLIC, ILC and the muon collider. Our conclusions are presented in the final section.

\section{Neutrino mass models and PMNS non-unitarity
\label{sec:models}}
\noindent\textbf{\emph{Type I seesaw.}\ }
In the type I seesaw model~\cite{Minkowski:1977sc,Yanagida:1979as,Mohapatra:1979ia,Gell-Mann:1979vob}, three heavy singlet Majorana fermions $N_R=(N_1,N_2,N_3)$ are added to the SM. Consequently, the 
matrices $\mM$ in \eq{eq:M} and $U$ in \eq{eq:U} have dimension $6\times 6$.\footnote{A 
 minimal version of the type I seesaw, consistent with all current neutrino data and predicting a massless lightest neutrino,  can be obtained by adding only two new neutral fermions~\cite{King:1999mb,King:2002nf,Frampton:2002qc}.}
In the basis $(\nu_L,N_R^c)$ 
with $ \nu_L =\left(\nu_e,\nu_\mu,\nu_\tau\right)$
the mass matrix has the following structure:
\begin{equation}
 \mM_{T1} = \begin{pmatrix}
      0 & m_D \\
      m_D^T & \hat M
  \end{pmatrix}   \;,
 \label{eq:MTI}
\end{equation}
where $\hat M$ is a matrix of Majorana masses that, with a proper choice of basis, can be taken as diagonal. The matrix $m_D$, proportional to the Higgs boson vacuum expectation value (vev), hosts the Dirac masses. The suppression  of the light neutrino masses is obtained under the natural assumption that all the entries in 
the $3\times 3$ matrix $\theta = m_D \hat M^{-1}$ are $\ll 1$. In this limit, the light neutrino mass matrix is given by
\begin{equation}
m_\nu = V_L^* \hat m_\nu V_L^\dagger \simeq - m_D \frac{1}{\hat M} m_D^T \,,
\label{eq:mnuT1}
\end{equation}
where $ \hat m_\nu= \diag(m_1,m_2,m_3)$ 
and $V_L$ is a unitary matrix. Note that $V_L$  differs from the exact (non-unitary) light neutrino mixing
matrix $U_\nu$ by terms of  $O(m_D^2/\hat{M}^2)$.
It is easily seen that, after neglecting the unphysical minus sign,  
the last expression  is reproduced by writing $m_D$ in the Casas-Ibarra (CI) 
parametrization~\cite{Casas:2001sr} as 
\begin{equation}
m_D = V_L^* \hat m_\nu^{\frac{1}{2}}  R^T \hat M^{\frac{1}{2}} \,,
\label{eq:mDTI}
\end{equation}
provided that $R$ is an arbitrary complex $3\times 3$ orthogonal matrix
and that the Yukawa couplings $Y$ determining the Dirac mass matrix $m_D = Y v$ are perturbative.\footnote{The matrix $R$  can be parametrised, for example,  as 
$R = \diag (\pm 1, \pm 1, \pm1) R^{(23)} (z_{23})R^{(13)}(z_{13})R^{(12)}(z_{12})$, with $R^{(ij)}$  being a rotation matrix acting in the $(ij)$ plane with a complex angle $z_{ij}$~\cite{Casas:2001sr,
Hambye:2003rt}.\label{footnote:3}}

At the leading order in $\theta $ 
the complete mixing matrix that diagonalises $\mM_{T1}$ in \eq{eq:MTI} is given by \cite{Ibarra:2011xn}
\begin{equation}
U \approx  \begin{pmatrix}
      1 - \frac{1}{2} \theta \theta^\dagger  & \theta \\
   -  \theta^\dagger & 1 -  \frac{1}{2} \theta^\dagger \theta
  \end{pmatrix}   \;, 
 \label{eq:UTI}
\end{equation}
and hence, in type I seesaw models, the deviation from unitarity in \eq{eq:diag} is given by  
\begin{equation}
\delta_\alpha = \sum_{i=4}^6 |\theta_{\alpha i}|^2 \,, 
\end{equation}
in which, by using \eq{eq:mDTI}, the light-heavy mixing matrix can be written as
\begin{equation}
\theta = V_L^* \sqrt{\hat m_\nu} \,  
R^T \frac{1}{\sqrt{\hat M}}   \, 
\label{eq:thetaTI}\,. 
\end{equation}
This shows that, in the type I seesaw model,  the active-sterile neutrino mixing is intrinsecally related to the 
suppression of  light neutrino masses.

Can some of the parameters $\delta_\alpha$ of the type I seesaw be
sufficiently large to be observable? In principle, yes. However,  taking as an order of magnitude estimate $\hat m_{\nu}\sim O(0.1\,\mathrm{eV})$ 
and $\hat M \sim O(400\,\mathrm{GeV})$ (for the consistency 
of the seesaw approximation, the lowest possible values for the entries in $\hat M$ 
cannot be below the electroweak scale), we see that  violation of unitarity 
at the level of $\delta_\alpha\sim 10^{-4}$   can be ensured only if  some entries in the matrix $R$ 
exceed $O(10^4)$. Clearly, this requirement implies a certain amount of  fine-tuning, in particular in the structure of the Dirac mass matrix $m_D$ (see \eq{eq:mDTI}).

\medskip

\subtopic{Inverse seesaw and double seesaw models.}
In the inverse seesaw (IS)~\cite{Mohapatra:1986bd}
and double seesaw (DS)~\cite{Mohapatra:1986aw}  
models, two 
triplets of SM singlet fermions $N_R=(N_1,N_2,N_3)$
and $S_L=(S_1,S_2,S_3)$ are introduced.
The mass matrix for the neutral states is $9\times 9$ and,
in the basis $\mN = \left( \nu_L, N_R^c, S_L\right) $, 
has the following structure:
\begin{equation}
 \mM_{IS,DS} = \begin{pmatrix}
      0 & m_D^T & 0 \\
      m_D & 0 & M  \\
          0&   M^T &  \hat\mu
  \end{pmatrix}  = \begin{pmatrix}
      0 & \mathbb{M}_D^T \\
      \mathbb{M}_D &  \mathbb{M} 
  \end{pmatrix}  \;, 
 \label{eq:MIS}
\end{equation}
where, in the second expression, 
  \quad 
  \begin{equation}
\mathbb{M}_D =\begin{pmatrix} m_D \\ 0\end{pmatrix},\quad 
\mathbb{M} =\begin{pmatrix} 0& M \\ M^T &\hat\mu\end{pmatrix}\,.
 \label{eq:MDM}
   \end{equation}
Without loss of generality, the Majorana mass term $\bar S_L^c \hat\mu S_L $ in 
Eqs.~(\ref{eq:MIS}) and (\ref{eq:MDM}) can be taken as diagonal by a field redefinition of the $S_L$ neutral fermions. The matrix $M$ that couples $N_R$ and $S_L$ can be factorised as $U_R \hat M W_L^\dagger$, with $\hat M$ diagonal with real non-negative entries. The unitary matrix  $U_R$ can then be absorbed in a redefinition of the $N_R$ fields, while the unitary matrix $W_L$ is defined in terms of three real and three imaginary physical parameters~\cite{Anamiati:2016uxp}.
 
The mass matrix  $ \mM_{IS,DS}$ can result in a suppression of the active neutrino masses if suitable hierarchies among its entries are assumed.
The IS model assumes  $\hat\mu \ll m_D\ll M$, so that an approximate diagonalization yields the light neutrino mass matrix
 \begin{equation}
     m_\nu = V_L^* \hat m_\nu V_L^\dagger \simeq  
   - \mathbb{M}_D^T \frac{1}{\mathbb{M}} \mathbb{M}_D 
     =
     m_D^T\frac{1}{M^T}\hat\mu \frac{1}{M} m_D
     \label{eq:mnuIS}\,.
 \end{equation}
This equation shows that a suppression of the light neutrino masses can be obtained through  small values of the lepton number violating (LNV) entries in  $\hat \mu$,  without the need for exceedingly small values of the ratio $m_D/M$. Consequently,  large values of the $N_R$ masses 
are not mandatory, and these new states 
could well lie within experimental reach. In contrast, the present analysis assumes that, in all cases, the heavy neutrino masses lie well above the energy scale of the experiments considered.
However, the IS model, as well as the DS and the linear seesaw (LS) models discussed below, allow the suppression of light neutrino masses to be effectively decoupled from the active-sterile neutrino mixing angle. As a result, in  these models the mixing can remain sizeable without requiring unnatural fine-tuning.

Adopting a CI-like parameterisation in the IS model, we  have: 
\begin{equation}
m_D =  M \frac{1}{\sqrt{\hat \mu}}\, R\, \sqrt{\hat m_\nu} \, V_L^\dagger\,.
\label{eq:mDIS}
\end{equation}
After diagonalization, the  $N_R$ and $S_L$ states give rise to a pair of heavy quasi-Dirac neutrinos $N_\pm$ with masses of $O(M)$ and a characteristic splitting of $O(\hat \mu)\ll O(M)$. The mixing between the light and heavy sectors is regulated by the block diagonal $1\times 2$ matrix
\begin{equation}
\theta = \mathbb{M}_D^T \frac{1}{\mathbb{M}}  =
\begin{pmatrix}
-m_D^T\frac{1}{M^T}\hat\mu \frac{1}{M}, & m_D^T \frac{1}{M^T} 
    \end{pmatrix} \,.
    \label{eq:thetaIS}
\end{equation}
Given that the ratio $\hat\mu \frac{1}{M}$ is a suppression factor within the IS model, the leading contribution to the PNMS unitarity violation in $\theta\theta^\dagger$ is given by the second block, that, with a slight abuse of notation, we denote $\theta_{\mathbf{12}}$. We have 
\begin{equation}
\theta_{\mathbf{12}} \simeq m_D^T \frac{1}{M^T} 
= V_L^* \sqrt{\hat m_\nu} R^T \frac{1}{\sqrt{\hat \mu}}\,,
\label{eq:theta12}
\end{equation}
which, notably, is not suppressed by the large scale $M$. 
Taking $\hat m_{\nu}\sim O(0.1\,\mathrm{eV})$ 
and $\hat \mu \sim O(100\,\mathrm{keV})$, we see that 
it would be sufficient to have entries of $O(10)$ in $R$ to 
generate a violation of unitarity at the level of $\delta_\alpha \sim 10^{-4}$.
Entries of this magnitude are not forbidden by symmetry 
arguments or by the requirement of Yukawa 
perturbativity, and various analysis 
of neutrino phenomenology based on $R$-matrix 
scans often extend to even larger 
values~\cite{Casas:2010wm}.\footnote{In 
$R$-matrix scans, the real part of the complex 
angles $z_{ij}$ (see footnote~\ref{footnote:3})
is generally varied as an ordinary angle, say 
$0\leq \Re[z_{ij}]\leq 2\pi$, while in 
principle $\Im[z_{ij}]$ can take values in 
the range $[-\infty,\infty]$. 
Of course too large values lead to non-perturbative Yukawa couplings, making the approach inconsistent,
regardless of naturalness considerations. 
Let us note, however, that taking 
$ \Im[z] \simeq \Re[z] \simeq \pi$,
well within their natural range, 
yields $|\cos z| = |\cosh \pi| = 11.6 $,
while the type I seesaw perturbative limit 
$|R_{ij}| \leq v^2/(M_i m_j)$~\cite{Casas:2010wm}
where $M_i\,(m_j)$ denote heavy (light) mass eigenvalues, can easily exceed $O(100)$. 
}
 
The DS model assumes for the entries in the mass matrix in \eq{eq:MIS}
the hierarchy  $\hat\mu \gg M \gg m_D$. The expression for the light neutrino mass matrix given in \eq{eq:mnuIS} and the CI-like parametrization in \eq{eq:mDIS} remain valid also in this case. However, in the DS model, $\hat\mu \frac{1}{M}$ represents an enhancement factor and so the leading term sourcing the PNMS unitarity violation in $\theta\theta^\dagger$ is the first block in \eq{eq:thetaIS}
\begin{equation}
\theta_{\mathbf{11}} = m_D^T \frac{1}{M^T} \hat \mu \frac{1}{M}
= V_L^* \sqrt{\hat m_\nu} R^T \sqrt{\hat \mu}
\frac{1}{M} \,,
\label{eq:theta11}
\end{equation}
which is enhanced by a factor $\sqrt{\hat \mu/M}\gg 1$ 
with respect to the type 1 seesaw case.

\medskip

\subtopic{Linear seesaw model.}
Another model that can be realised with new degrees of freedom at a low scale is the LS model~\cite{Akhmedov:1995ip,Akhmedov:1995vm}, which involves the same number of additional neutral fermions as in the previous case. 
In the basis $(\nu_L, N^C_R,S_L)$, the mass matrix
reads
\begin{equation}
 \mM_{LS} = \begin{pmatrix}
      0 & m_D^T & m_L^T \\
      m_D & 0 & M  \\
          m_L&   M^T & \hat\mu
  \end{pmatrix}   \;,
 \label{eq:MLS}
\end{equation}
where the non-zero 31 entries $m_L$ are 
generated by the vev of a Higgs doublet similarly to $m_D$ and, thus, are naturally of the order of the weak scale. Assuming the hierarchy  $m_L , m_D \ll M$, an   approximate diagonalization performed as in \eq{eq:MIS}, with 
$\mathbb{M}_D^T =(m_D^T, \   m_L^T)$, yields
\begin{equation}
    m_\nu \simeq m_D \frac{1}{M^T} \hat\mu \frac{1}{M} m_D -
    m_L^T \frac{1}{M} m_D  - m_D^T\frac{1}{M^T} m_L\,. 
   \label{eq:mnuLS}
\end{equation}
Note that if $\hat\mu \to 0$, the expression becomes linear in the 
active neutrino Yukawa couplings $Y=m_D/v$ to the right-handed neutrinos $N_R$,
hence the name {\em linear seesaw}.
The matrix
\begin{equation}
\theta = \mathbb{M}_D^T \frac{1}{\mathbb{M}}  =
\begin{pmatrix}
-m_D^T\frac{1}{M^T}\hat\mu \frac{1}{M} + m_L^T \frac{1}{M}, & m_D^T \frac{1}{M^T} 
    \end{pmatrix}
    \label{eq:thetaLS}
\end{equation}
controls the mixing between the light and heavy sectors of the theory. 

A CI-like parametrization can be written 
down also for the LS model~\cite{Herrero-Brocal:2025zpb,Garcia:2025frz}, however, it has a block matrix form that renders the expressions for the deviations from unitarity more complicated. From \eq{eq:thetaLS} we can see 
that the unitarity violation parameter given by $\theta\theta^\dagger$  
contains three additional terms besides the ones in \eq{eq:theta12}
and \eq{eq:theta11}, of which two are a pair of Hermitian conjugate combinations~\cite{Hettmansperger:2011bt}. All in all, the expected order of magnitude of unitarity violation is given by
\begin{equation}
    \theta \theta^\dagger \sim O\left[\frac{m_D^2}{M^2} \left(
    1+
    \frac{m_L^2}{m_D^2} + \frac{m_L}{m_D}\frac{\hat\mu}{M} +  
   \frac{\hat\mu^2}{M^2}\right)\right]\,.
\end{equation}
Which of these terms will dominate depends on the specific choice 
of the model parameters. Assuming, for example, that the terms containing $\hat\mu$ are negligible, the violation of unitarity is
determined by terms of order $(m_D^2+m_L^2)/M^2$, and can be sizable if $m_D$ or $m_L$ are not much smaller than $M$.

\section{Unitarity violation at lepton colliders}
\label{sec:cross-section}
\begin{figure}[t]
  \begin{center}
  \includegraphics[width=\linewidth]{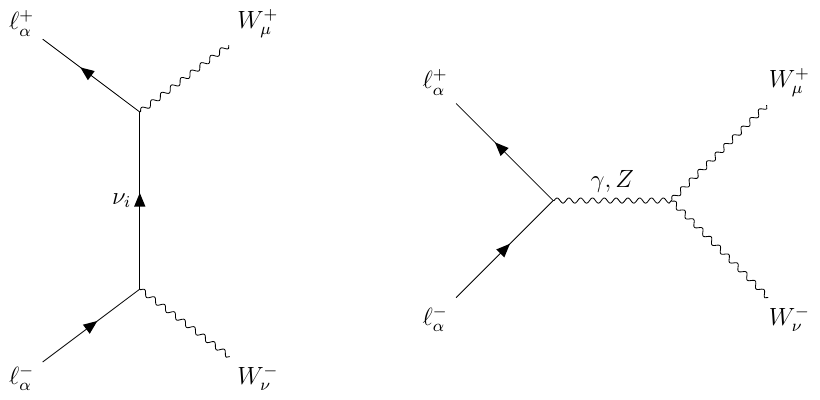}
  \caption{\small Tree level Feynman diagrams for the processes $\ell_\alpha^+\ell_\alpha^-\to W^+W^-$, $\alpha=e$, $\mu$, $\tau$. As explained in the text, the Higgs boson contribution is neglected.  The arrows on the fermion lines indicate the momentum flow. 
  \label{fig:diagrams} 
  } 
  \end{center}
  \end{figure}

\bigskip
\noindent
The violation of unitarity in the PMNS matrix is expected to break perturbative unitarity at the amplitude level in the process\footnote{The first computation of the related tree-level cross section can be found in Ref.~\cite{Alles:1976qv}. Radiative corrections to the process were first discussed in Refs.~\cite{Lemoine:1979pm,Philippe:1981up,Bohm:1987ck}.  }
\be
\ell^{+}_{\alpha}(p_1)\, \ell^{-}_{\alpha}(p_2) \rightarrow W^{+}(k_1)\, W^{-}(k_2)\, ,
\ee
where $\alpha = e,\mu,\tau$ label the initial state leptons. In the SM, this process proceeds via 
$s$-channels mediated by the Higgs boson, $Z$ boson, and photon, and via 
$t$-channel involving the exchange of the neutrino mass-eigenstates $\nu_{i}$. The corresponding tree-level Feynman diagrams are shown in Fig.~\ref{fig:diagrams}. In order to assess the power of lepton colliders to constrain non-unitarity effects, we restrict ourselves to consider electron and muon initial states, and work in the limit where the corresponding masses vanish---as justified by the typical energy of the processes we consider. In this limit we then neglect the Higgs boson $s$-channel contribution. 
We can then write the amplitude 
for  $\ell_\alpha^{+} \ell_\alpha^{-} \rightarrow W^{+} W^{-}$ as:
\be
      \mathcal{M}_{\alpha} =
     \mathcal{M}_{\alpha}^{t} + 
     \mathcal{M}_{\alpha}^{s}\,.
     \label{eq:amp_eeWW}
\ee
where $\mathcal{M}_{\alpha}^{t}$ and  $\mathcal{M}_{\alpha}^{s}$ represent the $s$-channel and  $t$-channel amplitudes, respectively. The potential presence in $\mathcal{M}_{\alpha}^{t}$ of terms spoiling the PMNS matrix unitarity prevents the exact cancellation of the terms proportional to the final state momenta contained in the two amplitudes, which gauge invariance otherwise ensures. 

In order to expose the origin of these terms, consider the $t$-channel contribution corresponding to diagram $(a)$ in Fig.~\ref{fig:diagrams}. We have 
\begin{equation}
\mathcal{M}_{\alpha}^{t} = \sum_{i=1}^{3} \left|{U^{\alpha i}_{\nu}}\right|^2 \mathcal{M}_i^{t}
\simeq \mathcal{M}_\nu^{t} \sum_{i=1}^{3} \left|{U^{\alpha i}_{\nu}}\right|^2\,,
\end{equation}
where the second equality holds once neutrino mass square differences
are neglected so that $\mathcal{M}_i^{t} = \mathcal{M}_\nu^{t}$ can be factored out
from the sum. 
If the PMNS mixing matrix is unitary, then 
\begin{equation}
\label{eq:3by3unitarity}
    \sum_{i=1}^{3} \left|{U^{\alpha i}_{\nu}}\right|^2 = 1\,,
\end{equation}
holds for each row $\alpha$. 

In frameworks where the SM neutral fermion sector is enlarged by adding 
$n$ new states that mix with the SM neutrinos, as in the neutrino mass models reviewed 
in Section~\ref{sec:models},  unitarity holds for the full 
 $(3+n)\times (3+n)$ matrix $U$. However, if the characteristic mass scale of the 
heavy neutrinos  lies well above the typical energy at which the $W$ boson pair production is probed, the $t$-channel diagrams involving their exchange are effectively negligible.
The sum over $t$-channel contributions is therefore restricted to the three light propagating degrees of freedom, whose mixing terms do not saturate the unitarity relation,  see \eq{eq:diag}. 
The quantities  $\delta_\alpha$ defined in~\eq{eq:diag} thus parametrize the deviation from unitarity for the  specific initial state lepton $\alpha = e,\mu, \tau$. As we will show below, this deviation spoils 
the cancellation of  $s$- and $t$-channel contributions, thereby resulting in an anomalous growth of the cross section as long as the energy remains 
below the mass scales of the new states.    

The amplitude in Eq.\eqref{eq:amp_eeWW} can be compactly written as
\bea
{\cal M}_{\alpha}&=&-ie^2 \Big[\bar{v}(p_1) \Gamma_{\alpha}^{\mu\nu} u(p_2)\Big]
\varepsilon^{\star}_{\mu}(k_1)\varepsilon^{\star}_{\nu}(k_2)
\, ,
\label{MDYWW}
\eea
where we indicated with $\varepsilon_{\mu}(k_1)$ and $\varepsilon_{\nu}(k_2)$ the polarization vectors of the $W^+$ and $W^-$, respectively. The effective vertex $\Gamma^{\mu\nu}_{\alpha}$ is given under the approximation of massless initial state by
\bea
\Gamma_{\alpha=e,\mu,\tau}^{\mu\nu}&=&\frac{1}{s}
\left(\gamma_{\beta} \bar{g}_V-\gamma_{\beta}\gamma_5 \bar{g}_A\right)
V^{\beta\nu\mu}(k_1+k_2,-k_2,-k_1)\nonumber\\
&+&\frac{\xi_{\alpha}}{4 t \ssW^2}\gamma^{\nu}\left(\slashed{p}_2-\slashed{k}_2\right)\gamma^{\mu}(1-\gamma_5)\,,
\label{Gamma}
\eea
with $\ssW= \sW$, $\theta_W$ being the Weinberg angle, $e$ the unit  electric charge, and 
\begin{equation}
  \xi_\alpha \equiv \sum_{i=1}^{3} |U_\nu^{\alpha i}|^2 = 1 - \delta_\alpha\,.
\end{equation}

The flavor universal effective couplings $\bar{g}_{V,A}$ in \eq{Gamma} are given by
\be
\bar{g}_V=-1+\frac{g_V\chi}{\ssW^2}\, ,~~
\bar{g}_A=\frac{g_A\chi}{\ssW^2}\, ,
~~ \chi=\frac{s}{2(s-\MZ^2)}\, ,
\label{effgva}
\ee
where  $\MZ$ is the $Z$ boson mass, $g_{V}=-\frac{1}{2}+2\ssW^2$,
$g_{A}=-\frac{1}{2}$. The $\chi$ term in \eq{effgva}, which weights the contribution of the virtual $Z$ 
channel, is real since we neglect the $Z$ width contribution. The function $V_{\beta\nu\mu}(k_1,k_2,k_3)$ arises from the Feynman rule for the trilinear vertex $V_{\beta}(k_1)~W^+_{\nu}(k_2)~W^-_{\mu}(k_3)$, with $V\in\{\gamma ,Z\}$, and is given by 
\bea
V_{\beta\nu\mu}(k_1,k_2,k_3)&=&(k_1-k_2)_{\mu}g_{\beta\nu}
+(k_2-k_3)_{\beta} g_{\mu \nu}\nonumber \\
&+&(k_3-k_1)_{\nu} g_{\beta\mu}\, ,
\label{V3}
\eea
for incoming momenta: $k_1+k_2+k_3=0$. The Mandelstam variables $s$ and $t$ appearing in the above equations are given by
\be
s=(p_1+p_2)^2, \quad t=(p_2-k_2)^2\,. 
\ee

The square amplitude for the process
$\ell_{\alpha}^+\,\ell_{\alpha}^- \to W^+W^-$, averaged over the initial state spins and summed over the final state polarizations, can be written as
\be 
|\xbar{{\cal M}}_{\alpha}|^{2}=|\xbar{{\cal M}}^{\SM}|^{2}+
\delta_{\alpha} ~\Delta_{1} +
\delta^2_{\alpha} ~\Delta_{2}\,,
\label{M2tot}
\ee
where $|\xbar{{\cal M}}_{\alpha}^{\SM}|^{2}$ is the SM contribution and the extra terms vanish in absence of unitarity violation. 

\begin{figure}
    \centering
    \includegraphics[width=\linewidth]{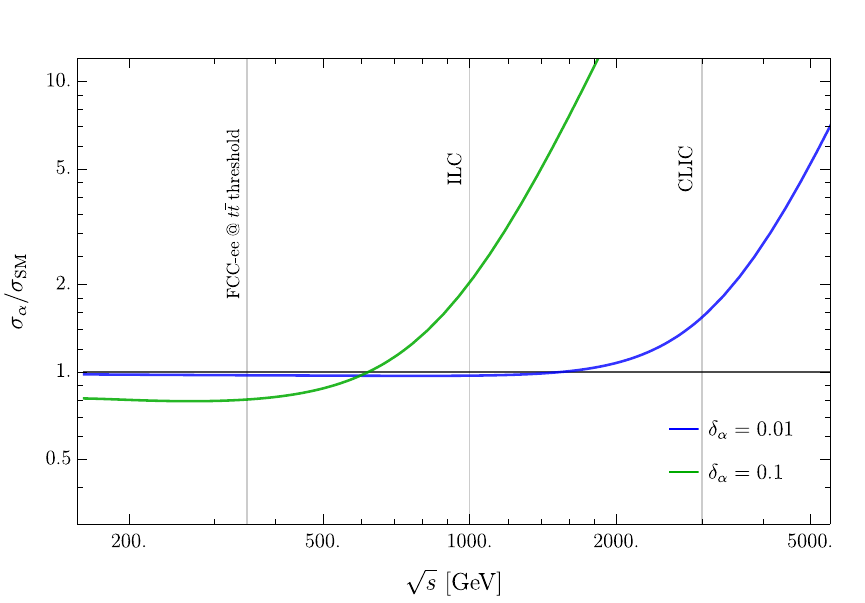}
    \caption{The ratio of the total cross section  including the NP contributions (\eq{eq:diff_xs}) for
     $\delta_\alpha = 0.01$ (blue solid line) and $\delta_\alpha = 0.1$ (green solid line), relative to the SM prediction ($\delta_\alpha = 0$) as a function of the center of mass energy.}
    \label{fig:sigma-ratio}
\end{figure}

By retaining only the leading orders in the $s/\MW^2 \gg 1$ expansion, with $\MW$ the $W$ boson mass, we find
\bea
\Delta_{1}\!\!&\simeq&\!\!
- \frac{\alpha_{\W}^2\pi^2}{2}\left(\frac{s}{\MW^2} \right)
\St^2\left[ 2-\frac{\MZ^2}{\MW^2}\left(1-2\ssW^2\right) \right]\,,
\nonumber\\
   \Delta_{2}\!\!&\simeq&\!\!
   \frac{\alpha^2_{\W}\pi^2}{4}
   \Big[\!
     \left(\frac{s^2}{\MW^4}\right)\St^2+
        \left(\frac{s}{\MW^2}\right)4\left(3+\Ct^2\right) \!\Big]\!, ~~~~
   \label{DeltaM}
   \eea
 with $\alpha_{\W}= e^2/(4\pi\ssW^2)$ and $s^2_W$ denoting the sine square of the electroweak angle, while $\Ct = \cos \Theta$ and $\St=\sin\Theta$, with $\Theta$ being the scattering angle between the $e^+$ and $W^+$ momenta.
Notice that the correction $\delta_\alpha\, \Delta_{1} \propto - \delta_\alpha s$,  is always negative, 
while $\delta_\alpha^2\, \Delta_{2}$, that contains a term $\propto \delta_\alpha^2 s^2$,  is always positive.
Thus, for small $\delta_\alpha$,  the unitarity violation effects 
will suppress the cross section  at low energies with respect to the SM. However, for any 
given value of $\delta_\alpha$ there will be a certain value of the center-of-mass energy $s$ for which the $\delta_\alpha^2 \Delta \xbar{{\cal M}}_{2}$ correction 
starts dominating, and the cross section will then overshoots the SM prediction.
This behavior is clearly illustrated in Fig.~\ref{fig:sigma-ratio}.
For the sake of the present analysis, we have used the exact expressions of the $\Delta\xbar{{\cal M}}_{1}$ and $\Delta\xbar{{\cal M}}_{2}$ terms, which are given in Appendix~\ref{sec:Appendix-A}.

The differential cross section for the process  reads:
\begin{equation}
    \left(\frac{d \sigma_\alpha}{d \Omega}\right) = \frac{\beta_{\W} }{64 \pi^2 s}|\xbar{{\cal M}}_{\alpha}|^{2}\,,
    \label{eq:diff_xs}
\end{equation}
where $\beta_{\W}= \sqrt{1 - 4\MW^2/s}$ is the $W$ boson velocity in the center of mass frame.

In Fig.~\ref{fig:sigma-ratio} we plot the ratio of the total cross section \eq{eq:diff_xs} for the two values $\delta_\alpha =0.1$ and $\delta_\alpha=0.01$,  
relative to the SM prediction (corresponding to $\delta_\alpha=0$), as a function of $\sqrt{s}$. If a deviation from unity is measured for this ratio,  this 
could be straightforwardly interpreted as the effect of unitarity-breaking in the lepton mixing matrix. As the center-of-mass energy reaches the mass scale of these new states, the cross section starts to fall with increasing energy. In the following sections, we exploit the deviations induced by the $\delta_\alpha$ parameters to assess the sensitivity of collider experiments to the unitarity violation effects they imply.

\section{Lepton collider bounds on $\delta_{\alpha}$}
\label{sec:bounds}

In order to investigate the power of lepton colliders to constrain the diagonal unitarity violating effects encoded in the $\delta_\alpha$ parameters, we set up $\chi^2$ tests that highlight the parameter ranges allowed at $95\%$ confidence level (CL). 

Starting with LEPII, we utilise the measurements~\cite{Behner:2000yt} reported in Tab.~\ref{tab:LEP:table} of the $e^+e^-\to W^+ W^-$ cross section, $\sigma_i$, and the relative experimental errors, $\epsilon_i$, where the subscript $i=1,2,\dots 6$  denotes the six different center of mass energies $\sqrt{s_i}$ at which the measurements were performed.
\begin{table}[h]
\centering
\begin{tabular}{c | c | c}
\toprule
$\sqrt{s_i}\;[\mathrm{GeV}]$ &
$\sigma_i\;[\mathrm{pb}]$ &
$\epsilon_i\;[\mathrm{pb}]$ \\
\midrule
183 & 15.83 & 0.36 \\
189 & 16.05 & 0.22 \\
192 & 16.80 & 0.50 \\
196 & 17.39 & 0.35 \\
200 & 16.91 & 0.32 \\
202 & 17.11 & 0.46 \\
\bottomrule
\end{tabular}
\caption{LEPII data used for the $\chi^2$ analysis with six degrees of freedom.}
\label{tab:LEP:table}
\end{table}

The $\chi^2$ test we use is then given by
\begin{equation}
    \chi^2 = \sum_{i=1}^6\left(\frac{\sigma_i - \frac{\sigma_i}{\sigma_{\rm SM}(s_i)} \sigma_e(s_i)}{\epsilon_i}\right)^2\leq 12.592\,,
\end{equation}
where $\sigma_e$ is the tree-level cross section for the 
 $e^+e^-\to W^+ W^-$ process, including effects of violation of 
 PMNS unitarity ($\delta_e\neq0$), 
 $\sigma_{\rm SM} \equiv \sigma_e(\delta_e =0)$
is the SM cross section, 
 and  $\sigma_i/\sigma_{\rm SM}(s_i)$
 is a rescaling factor that effectively accounts for next-to-leading order (NLO) corrections, experimental cuts, and detector efficiencies. The resulting bound on  $\delta_e$ that accounts for unitarity violations in  the first row of the PMNS matrix is
\begin{equation}
    \delta_e \lesssim 0.0135\quad(\text{LEPII}),
\end{equation}
at $95 \%$\,CL. 

To forecast a possible bound obtainable at the FCC-ee, we perform a $\chi^2$-test targeting the projected number of events expected at the center of mass energy of $\sqrt s = 350 $ GeV  with a benchmark luminosity of $\mathcal{L}=1.8\, {\rm ab}^{-1}$, corresponding to 5 years of operation~\cite{FCC:2018evy,FCC:2025lpp}. We obtain:
\begin{equation}
  \chi^2 = \frac{\left(N_{\rm obs} - N_{\rm exp}\right)^2}{N_{\rm exp}} = \mathcal{L}\frac{\left(\sigma_e - \sigma_{\rm SM}\right)^2}{\sigma_{\rm SM}}\leq 3.841\,,
  \label{chi2-FCCee}
\end{equation}
where we have taken into account only the statistical error. 
Systematic errors related to the setup of the experiment, as well as the contribution 
to the overall uncertainty of yet unknown higher order corrections are expected to 
be important, although one can hope that, by the time the measurement will be performed, 
these uncertainties  will eventually be reduced at a level not exceeding the statistical error.

With this assumption, we obtain for the  95\% CL limit achievable at the FCC-ee, the following projection:
\be
\delta_{e} \lesssim 1.6 \times 10^{-4}\quad (\text{FCC-ee})\,.
\ee
We remark that even if the effects of PMNS unitarity violation
on the cross section generally grow with the energy, the FCC-ee bound is primarily 
due to  the extremely high luminosity of the machine, which enhances the sensitivity of the process to the anomalous $\delta_e$ effect, even for a center of mass energy not far above the electroweak scale. 
Future $e^+e^-$ linear colliders and/or muon colliders,  which are  designed to run at center of mass energies well above the TeV scale, will be able to probe genuine high-energy effects due to, respectively,  $\delta_e$ and $\delta_\mu$. In particular, we consider the cases of the future $e^+e^-$ International Linear Collider (ILC) \cite{ILC:2013jhg,Bambade:2019fyw,ILCInternationalDevelopmentTeam:2022izu} and Compact Linear Collider
  (CLIC)  \cite{Linssen:2012hp,CLIC:2018fvx,Brunner:2022usy}, respectively expected to run at 1 TeV for an integrated luminosity of $8{\rm ab}^{-1}$, and at 3 TeV for an integrated luminosity of $5{\rm ab}^{-1}$. Repeating the previous analysis we then find
\bea
\delta_{e} &\lesssim& 9.6\times 10^{-5} \quad (\text{ILC})\,, \nonumber\\
\delta_{e} &\lesssim& 9.1\times 10^{-5} \quad (\text{CLIC})\, ,
\label{delta-ILC-CLIC}
\eea
at a $95\%$ CL. For the muon collider, we consider two possible benchmark points: BM1 with $\sqrt{s} = 3\, {\rm TeV}$ and $\mathcal{L} = 1\, {\rm ab}^{-1}$, and BM2 with $\sqrt{s} = 10\, {\rm TeV}$ and $\mathcal{L} = 10\, {\rm ab}^{-1}$ \cite{InternationalMuonCollider:2024jyv}. In these cases, assuming as before that statistical errors dominate, we obtain at $95\%$ CL:
\bea
 \delta_{\mu} &\lesssim& 2.2 \times 10^{-4}\, ~~~ {\rm (BM1)} 
 \,,
 \nonumber\\
 \delta_{\mu} &\lesssim& 3.1\times 10^{-5}\, ~~~ {\rm (BM2)} 
 \, .
 \label{delta-muon}
\eea

\begin{figure}[h!]
    \centering
    \includegraphics[width=1\linewidth]{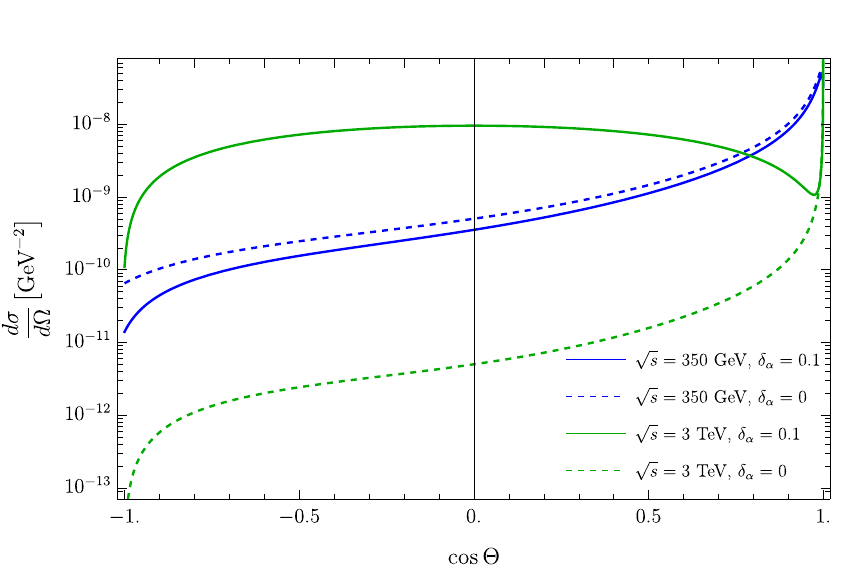}
    \caption{The angular distribution of the differential cross section at two fixed energies: $\sqrt{s} = 350 \text{ GeV}$  (blue solid line: including the NP contribution with $\delta_\alpha = 0.1$, blue dashed line: SM result) 
        and $\sqrt{s} = 3 \text{ TeV}$, corresponding to the center of mass energies of FCC-ee and CLIC  (green solid line: including  the NP contribution
        with $\delta_\alpha = 0.1$, green dashed line: SM result).  The different behaviour of the curves reflects the interplay between the two contributions appearing in Eq.~\eqref{M2tot} as a function of
          $\sqrt{s}$.
    \label{fig:angulardist}}
\end{figure}

To conclude the section, we mention the possibility of using angular cuts to isolate the new physics contribution from the SM background at a fixed energy, a strategy 
that is described in ref.~\cite{Gabrielli:2024bjw}, where an analogous study of the effects of 
violation of unitarity in the the quark mixing matrix is presented. The effectiveness of 
this strategy is illustrated in Fig.~\ref{fig:angulardist} for the benchmarks of FCC-ee and CLIC discussed above. Evidently, in the high energy regime where the cross section is dominated by the unitarity-violating terms, an angular cut centered around the scattering angle $\Theta = \pi/2$ 
would enhance the effects of PMNS unitarity violation, so that deviations from the SM prediction are more easily detected in this angular region.

\section{Unitarity tests at hadron colliders}

\begin{figure}[t!]
  \begin{center}
  \includegraphics[height=3.5 cm]{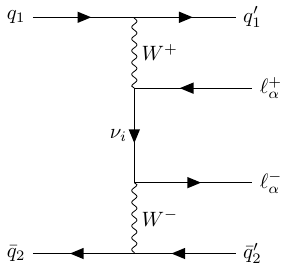}
  \includegraphics[height=3.5 cm]{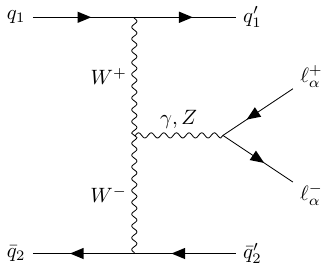}
  \caption{\small Tree level Feynman diagrams for the processes $p p\to jj\, \ell_\alpha^+ \ell_\alpha^-$, $(\alpha=e,\,\mu,\,\tau)$. The arrows on the fermion lines indicate the momentum flow. 
  \label{fig:diagram_h} 
  }
\end{center}
\end{figure}

Present and future hadron colliders can use the same method to probe the unitarity of the PMNS matrix by means of the inverse process
\begin{equation}
  p p \to \ell^+_\alpha\, \ell^-_\alpha\, j\, j \,,\qquad\alpha=e,\, \mu, \,\tau\,,
\end{equation}
proceeding via a pair of $W$ bosons produced from initial quarks, where on top of the charged lepton pair, two jets $jj$ appear in the final state. 
The corresponding (partonic) Feynman diagrams are shown in Fig.~\ref{fig:diagram_h}.

To estimate the size of the cross section for this process, we resort to the Effective Vector Boson Approximation (EVBA) \cite{Dawson:1986tc}.
This computational scheme replaces the virtual $W$ bosons emitted from the initial quark lines with real $W$ vector bosons, emitted with probabilities that depend on whether they are longitudinally (L) or transversely (T) polarized. The related partonic square amplitude admits the same decomposition presented in \eq{M2tot} for leptonic initial states, but we need to distinguish among the initial $WW$ polarizations. In particular, for the
$W^+W^-\to \ell_\alpha^+ \ell_\alpha^-$ process the square amplitude can be decomposed as
\bea
|M^{\AB}_{\WW}|^2&=&|M^{\AB}_{\rm SM}|^2 +\delta_{\alpha} \Delta^{\AB}_{1}
+\delta_{\alpha}^2 \Delta^{\AB}_{2}
\eea

where $AB=\{LL,TT,LT\}$ indicate the different polarization combinations and $|M^{\AB}_{SM}|^2$ denotes the SM contribution \cite{Dawson:1986tc,Dicus:1986az}. The expressions for the $\Delta M^{AB}_{1,2}$ terms are given in Appendix~\ref{sec:Appendix-B}.

Using the results given in Ref.~\cite{Green:2003if}, we obtain for the differential distribution of the cross section as a function of invariant mass  $m_{\ell \ell}$ of the resulting lepton pair:
\begin{widetext}
\begin{equation}
\frac{d \sigma}{d m_{\ell \ell}}(pp\to \ell^+\ell^-jj) =
\int_{m_{\ell \ell}^2/S}^1\frac{d a} 
{\sqrt{a}} \sum_{q_1,q_2} 
L^{q_1 q_2}(z)~ \Big[
L_{\LL}(a)~ \sigma^{\LL}_{\WW}(m_{\ell \ell}^2)
+L_{\TT}(a)~ \sigma^{\TT}_{\WW}(m_{\ell \ell}^2)+L_{\LT}(a
)~ \sigma^{\LT}_{\WW}(m_{\ell \ell}^2)\Big],\label{eq:EVBA}
\end{equation}
\end{widetext}
where  
$L_{AB}$ are the $WW$ luminosities for the $AB=(LL),\,(TT),\,(LT)$ polarizations, $\sigma^{AB}_{\WW}(m_{\ell \ell})$ are the corresponding polarized  cross sections for the $W^+W^-\to \ell^+\ell^-$ process evaluated at the lepton-pair invariant mass, and  the symbol $L^{q_1 q_2}(z)$  denotes
the differential parton luminosity for initial $q_1 q_2$ quark state carrying 
a fraction  $z =  m^2_{\ell\ell}/(a\,S)$ of the total hadronic center-of-mass energy square denoted by $S$, while the sum extends over all pairs of  $u$, $d$, $s$, $c$, $b$ quarks and antiquarks that can produce a $W^{+}W^-$ pair.\footnote{The contributions of the $c$ and $b$-quarks are strongly suppressed for the center of mass energy of the HL-LHC and, therefore, can be neglected in this setting. They must however be included when assessing the power of the upcoming FCC-hh  experiment.}

In the limit where the initial quark energy is much above the $W$ boson mass, the polarized $WW$ luminosities ~\cite{Dawson:1986tc} and the quark luminosities are given by 
\begin{widetext}

\bea
L_{\LL}(a)&=&-\left(\frac{\alpha_{\W}}{4\pi}\right)^2\frac{1}{a}
\left[(1+a)\log{a}+2(1-a)\right]\, ,
\nonumber\\
L_{\TT}(a)&=&-\left(\frac{\alpha_{\W}}{8\pi}\right)^2\frac{1}{a}
\left[2(1-a)(3+a)+(2+a)^2\log{a}\right]
\left[\log{\left( 
\frac{m^2_{\ell\ell}}{a\, M^2_W} 
\right)}\right]^2\, ,
\nonumber \\
L_{\LT}(a
)&=&\left(\frac{\alpha_{\W}}{8\pi}\right)^2\frac{1}{a}
\left[-7+6a+a^2-4(1+a)\log{a}\right]
\log{\left(\frac{m^2_{\ell\ell}}{a\,M_W^2}\right)}\,,
\nonumber \\
L^{q_1 q_2}(z)&=& 
4 \sqrt{\frac{z}{S}}
\int_{z}^{1} \frac{{\rm d}x}{x} f_{q_1}(x) f_{q_2}\left(\frac{z}{x}\right) \,,
\eea
\end{widetext}
where, with a slight abuse of notation, $L_{LT}$ indicates the sum of both the $L_{LT}$ and $L_{TL}$ contributions, and $f_{q_1}$ and $f_{q_2}$ denote the parton distribution functions of quarks $q_1$ and $q_2$ inside the colliding proton pairs.

In order to assess the possibility offered by hadron colliders, in the following we again compare the cross section obtained for a non-vanishing value of $\delta_\alpha$ to the SM case.

With respect to the cleaner case offered by lepton colliders, we expect that the presence of sizable backgrounds due to the $Z$ boson or the photon, for instance, will inevitably worsen the reach of our methodology. Nevertheless, we point out that these machines offer a unique possibility for testing unitarity breaking effects involving $\tau$ lepton pairs, which complement the results obtained at a lepton colliders using electron or muon initial states. Having said that, in this first exploratory analysis we  neglect the possible effects of  certain background processes in gauging the reach of the upcoming HL phase of LHC~\cite{HL-LHC-TDR,LHC-Design} and the FCC-hh collider,
as for example the contribution of diagrams similar to the first 
one in fig.~\ref{fig:diagram_h} with the $W$ bosons replaced by $Z$ bosons 
or photons, which would add incoherently to the $jj\ell^+\ell^-$ 
production. In first approximation this is justified since the 
contribution from the $Z$-diagrams is suppressed with respect to the $W^+W^-$ signal by the smaller $Z\bar qq$ and $Z\bar\ell\ell$ couplings~\cite{Dicus:1986az}, while the 
contribution from $\gamma$ diagrams can be kept under control by selecting final states containing jets with sizable transverse momentum~\cite{Cox:2005if}. However, more refined estimates should not dispense with a dedicated treatment of these, as well as other, background processes, and should also account
for detector effects specific to each experiment.

In order to derive the potential unitarity bounds that could be set by the high luminosity phase of LHC on $\delta_{\ell}$, we perform a $\chi^2$ test as in \eq{chi2-FCCee}, by considering the total cross section from \eq{eq:EVBA} integrated in the range
$200~{\rm GeV} \le \mWW \le 1~{\rm TeV}$. With the PDF4LHC21 PDFs given in Ref.~\cite{PDF4LHCWorkingGroup:2022cjn} and setting $\sqrt{S} = 13~{\rm TeV}$, the HL-LHC luminosity benchmark of
$\mathcal{L}=3~{\rm ab}^{-1}$~\cite{HL-LHC-TDR} gives
\be
\delta_{\alpha}< 1.0\times 10^{-3}~~~ (\text{HL-LHC, 95\% CL})\, ,
\ee
for every flavor $\alpha=e, \,\mu,\,\tau$. Notice that the bound is flavor blind, being obtained in the massless-fermion limit and under the assumption that the background is given solely by the SM process $W^+W^-\to \ell_\alpha^+\ell_\alpha^-$. As mentioned before, this constraint can thus also be applied to the $\delta_{\tau}$ case, which is not accessible at lepton colliders.

We extend the analysis to assess the reach of the proposed method at the future circular hadron collider FCC-hh~\cite{Mangano:2017tke,FCC:2018byv}, operating at $\sqrt{S} = 100~{\rm TeV}$. For the computation in \eq{eq:EVBA} we again use the PDF4LHC21 PDFs given in Ref.~\cite{PDF4LHCWorkingGroup:2022cjn}, with $\sqrt{S} = 100~{\rm TeV}$ and factorisation scale $\hat s = z S$. Using the $\chi^2$ test as in \eq{chi2-FCCee}, and considering the total cross section from \eq{eq:EVBA} integrated over the range $200~{\rm GeV} \le \mWW \le 9~{\rm TeV}$, we obtain
\be
\delta_{\alpha}< 4.4 \times 10^{-5} ~~~ (\text{FCC-hh, 95\% CL})\,
\ee
for every lepton flavor $\alpha =e, \,\mu,\,\tau$, and ignoring, again, the effect of potential backgrounds. 

\section{Conclusions}

We have proposed a strategy to probe the unitarity of the PMNS matrix using $W$ boson pair production at collider experiments. At electron-positron and muon colliders, the key observable is the ratio of the $\ell^+_\alpha\ell^-_\alpha \to W^+W^-$  ($\alpha =e,\,\mu$) 
cross section to its SM prediction: any non-zero value of the flavor-diagonal non-unitarity parameters $\delta_\alpha$ spoils the gauge cancellation between $s$- and $t$-channel amplitudes, inducing an anomalous energy growth in the cross section at energies below the mass threshold of the new heavy states responsible for the loss of unitarity in the mixing matrix. At hadron colliders, the same effect can be investigated by analyzing the inverse process, $p p \to W^+ W^- \to \ell^+_\alpha\ell^-_\alpha jj$  ($\alpha =e,\,\mu,\,\tau$). Thus,   
our strategy enables hadron colliders to probe possible unitarity-violating effects in the 
$\tau$ sector, which are inaccessible at experimental facilities based on electron or muon initial states. 

The results obtained for the flavor-diagonal non-unitarity parameters at a 95\% confidence level are summarized in Table~\ref{tab:future colliders}.  We remark that the applicability of our method is set by the mass of the
lightest new state $N$. For the lepton-collider projections this means $M_N \gg \sqrt{s}$, ranging from $M_N \gg 350$~GeV (FCC-ee benchmark) to $M_N \gg 10$~TeV (muon collider
BM2). For the hadron-collider bounds, instead, the relevant constraint is $M_N \gg m_{\ell\ell}^{\rm max}$, which is typically of the TeV order at HL-LHC and about 10 TeV at the FCC-hh.

Taken at face value, the bounds in Table~\ref{tab:future colliders} show that the proposed method can indeed complement, and potentially even outclass, the results of electroweak precision measurements especially in the electron and tau sectors. The extent to which loop corrections, background processes and detector effects will affect the proposed method will require a case-by-case dedicated study, which goes beyond the scope of this paper. 

\begin{table}[t!!]
\centering
\begin{tabular}{ c | c | c | c}
\hline
 Experiment & $\sqrt{s}\;[\mathrm{TeV}]$ &
$\mathcal{L}\;[\mathrm{ab^{-1}}]$ &
Bound \\
\hline
FCC-ee & 0.35 & 1.8 & $\delta_e \lesssim 1.6 \times 10^{-4}$\\
ILC & 1 & 8 & $\delta_e \lesssim 9.6 \times 10^{-5}$\\
CLIC & 3 & 5 & $\delta_e \lesssim 9.1 \times 10^{-5}$ \\
$\mu$-collider BM1 & 3  & 1 & $\delta_\mu \lesssim 2.2 \times 10^{-4}$ \\
$\mu$-collider BM2 & 10 & 10 & $\delta_\mu \lesssim 3.1 \times 10^{-5}$\\
HL-LHC & 13 & 3 & $\delta_{\alpha} \lesssim 1.0  \times 10^{-3}$ \\
FCC-hh & 100 & 30  & $\delta_{\alpha} \lesssim 4.4  \times 10^{-5}$ \\
\hline
\end{tabular}
\caption{Summary of the bounds achievable by the different collider experiments. The bounds obtainable at hadron collider hold for all flavors $\alpha=e,\,\mu, \,\tau$. }
\label{tab:future colliders}
\end{table}

\subsection*{Acknowledgments}
\noindent
It is a great pleasure to thank Emanuele Bagnaschi and Gennaro Corcella for insightful discussions in the early stages of this project. This work was supported by the Estonian Research Council grants TARISTU24-TK10, TARISTU24-TK3, RVTT3, and by the CoE TK 202 ``Foundations of the Universe’', and by the CERN Science Consortium of Estonia, grant RVTT3. The work of K.M. was supported by the Estonian Research Council grant PUTJD1256. The work of E.N. and L.M. was also supported by the 
Estonian Research Council team grant PRG1884. 

\vfill\clearpage

\begin{widetext}

\begin{appendix}
\section{}
  \label{sec:Appendix-A}
We provide below the complete analytical expressions for the quantities 
$\Delta_{1,2}$ appearing in \eq{M2tot}:
\bea 
\Delta_1&=&
\frac{4\alpha_{\W}^2\pi^2}{\left(1-\betaW^2\right)^2\left(1-2\betaW\Ct+\betaW^2\right)^2\left(\MZ^2-s\right)}
\Bigg\{s \left(1 - \betaW^2\right) \Big[-8 + 24 \betaW \Ct 
\nonumber\\
&-& \betaW^2 \left(7 + 9 \Ct^2\right) - 16 \betaW^3 \Ct
+ 2 \betaW^4 \left(7 + 3 \Ct^2 + 2 \Ct^4\right)
- 8 \betaW^5 \Ct^3 -  3 \betaW^6 \left(1 - \Ct^2\right)
\Big] \nonumber\\
&+& 2 \MZ^2 \Big[4 - 8 \betaW \Ct \left(2 -\ssW^2\right)+ \betaW^2 \Big(9 + \Ct^2 \left(11 - 13 \ssW^2\right) - 19 \ssW^2\Big)
  \nonumber\\
&+& 4 \betaW^3 \Ct \Big(-1 + 12 \ssW^2 + \Ct^2 \left(1 - 2 \ssW^2\right)\Big)
  - \betaW^4 \Big(10 + \ssW^2 + 3 \Ct^2 \left(2 + \ssW^2\right) +4\Ct^4 \left(1 -  \ssW^2\right)\Big)
\nonumber\\
&+& 4 \betaW^5 \Ct \left(3 + \Ct^2 - 10 \ssW^2\right) 
+ \betaW^6 \Big(1 + 15 \ssW^2 + 4 \Ct^4 \ssW^2 - \Ct^2 \left(5 - 13 \ssW^2\right)\Big)\nonumber\\
&-& 8 \betaW^7 \Ct^3 \ssW^2 - 3 \betaW^8 \left(1 - \Ct^2\right) \ssW^2
\Big]
\Bigg\}\,,
\label{eq:A1}
\\
\Delta_2&=&
\frac{4\alpha_{\W}^2\pi^2}{ \left(1-\betaW^2\right)^2\left(1-2\betaW\Ct+\betaW^2\right)^2}
\Big[4 - 16 \betaW \Ct + \betaW^2 (9 + 11 \Ct^2) - 4 \betaW^3 \Ct (1 - \Ct^2)
\nonumber\\
&-&  2 \betaW^4 (5 + 3 \Ct^2 + 2 \Ct^4)
+ 4 \betaW^5 \Ct (3 + \Ct^2) + \betaW^6 (1 - 5 \Ct^2)
\Big]\,.
\label{eq:A2} 
\eea
The definitions of the symbols appearing in \eqs{eq:A1}{eq:A2} can be found in section \ref{sec:cross-section}.
\section{}
\label{sec:Appendix-B}
We list below the expressions for the unitarity-violation corrections to the polarized square amplitudes $|M^{\AB}_{1,2}|^2$  for the process $W^+W^-\to \ell^+ \ell^-$, for the  combinations of polarization $AB=\{LL,TT,LT\}$:

\bea
\Delta^{\LL}_{1}&=&\frac{16\alpha_{\W}^2\pi^2
\left(1-\Ct^2\right)
  \left(\betaW^3-3\betaW+2\Ct\right)
  }{\left(1-\betaW^2\right)^2
  \left(1+\betaW^2-2\betaW\Ct\right)^2
  \left(s-\MZ^2\right)} \Bigg\{s\Big[
\left(1 - \betaW^2\right) \left(-3 \betaW + \betaW^3 + 4\Ct -
2 \betaW^2 \Ct\right)\Big]\nonumber\\
&-&2\MZ^2
\Big[2 \Ct - 3 \betaW (1 - \ssW^2) - \betaW^5 \ssW^2 - 6 \betaW^2 \Ct \ssW^2 + 
   2 \betaW^4 \Ct \ssW^2 + \betaW^3 (1 + 2 \ssW^2)\Big]
 \Bigg\}\,,
 \nonumber\\
\Delta^{\LL}_{2}&=&\frac{16\alpha_{\W}^2\pi^2  \left(1-\Ct^2\right)
  \left(\betaW^3-3\betaW+2\Ct\right)^2}
       {\left(1-\betaW^2\right)^2\left(1+\betaW^2  -2\betaW\Ct\right)^2}\,,
\nonumber\\
\Delta^{\TT}_{1}&=&\frac{32\alpha_{\W}^2\pi^2
    \left(1-\Ct^2\right)}{\left(1+\betaW^2-2\betaW\Ct\right)^2
    \left(s-\MZ^2\right)}   
  \Bigg\{
  s \Big[2-\betaW^4-3\betaW\Ct+3\betaW^3\Ct+4\Ct^2
  +\betaW^2\left(1-2\Ct^2\right)\Big]
  \nonumber\\
&-&  2 \MZ^2\Big[1 + 2 \Ct^2 - \betaW \Ct (2 - \ssW^2) - \beta^4 \ssW^2
  + 3 \betaW^3 \Ct \ssW^2 +  \betaW^2 (1 - \ssW^2 - 2 \Ct^2 \ssW^2)\Big]
  \Bigg\}\,,
  \nonumber\\
  \Delta^{\TT}_{2}&=&
  \frac{32\alpha_{\W}^2\pi^2
  \left(1-\Ct^2\right)
    \left(1+\betaW^2 -2\betaW\Ct+2\Ct^2\right)
}
       {\left(1+\betaW^2  -2\betaW\Ct\right)^2}\,,
       \nonumber\\
\Delta^{\LT}_{1}&=&-\frac{64\alpha_{\W}^2\pi^2}
 {\left(1-\betaW^2\right)\left(1+\betaW^2-2\betaW\Ct\right)^2
         \left(s-\MZ^2\right)}   
\Bigg\{s \left(1 - \betaW^2\right)
\Big[-1 + 2 \betaW^3 \Ct + 3 \Ct^2 + 6 \betaW \Ct^3 - 4 \Ct^4 - 
\betaW^2 (1 + 5 \Ct^2)\Big]
\nonumber\\
&+&\MZ^2 \Big[1 - 3 \Ct^2 + 4 \Ct^4 + \betaW^4 \left(1 + \Ct^2 (1 - 12 \ssW^2)
- 4 \ssW^2\right) + 
4 \betaW \Ct^3 (\ssW^2-2) + 4 \betaW^5 \Ct \ssW^2
\nonumber\\
&+& 
  4 \betaW^3 \Ct \left(3 (1 + \Ct^2) \ssW^2-2\right) + 
  2 \betaW^2 \big(1 - 2 \ssW^2 - 4 \Ct^4 \ssW^2 + \Ct^2 (5 - 2 \ssW^2)\big)
  \Big]
\Bigg\}\,,
\nonumber\\
  \Delta^{\LT}_{2}&=&
  \frac{32\alpha_{\W}^2\pi^2}{\left(1-\betaW^2\right)
    \left(1+\betaW^2  -2\betaW\Ct\right)^2}
  \Bigg\{
1 - 8 \betaW^3 \Ct - 3 \Ct^2 - 8 \betaW \Ct^3 + 4 \Ct^4 + 
  \betaW^4 \left(1 + \Ct^2\right) + 2 \betaW^2 \left(1 + 5 \Ct^2\right)
\Bigg\}\,.
\eea

\end{appendix}

\end{widetext}

\bibliographystyle{utphys}
\bibliography{Neutrino-unitarity.bib}

\end{document}